\documentstyle[osa,manuscript]{revtex}

\begin{document}
\title{Heisenberg model in a random field: phase diagram and tricritical behavior}
\author{Douglas F. de Albuquerque$^{\text{1,}}$ 
\footnote { {\bf E-mails addresses:} douglas@ufs.br, clavier@ig.com.br, spunki@bol.com.br}%
and Alberto S. de Arruda$^{\text{2}}$}
\address{$^1$Departamento de Matem\'{a}tica, Universidade Federal de Sergipe,
49100-000, S\~{a}o Cristov\~{a}o, SE, Brazil.%
\newline%
$^2$Departamento de F\'{\i }sica, Universidade Federal de Mato Grosso,
78060-900, Cuiab\'{a}, MT, Brazil.}
\date{
\today%
}
\maketitle

\begin{abstract}
By using the differential operator technique and the effective field theory
scheme we study the tricritical behavior of Heisenberg classical model of
spin-1/2 in a random field. The phase diagram in the {\it T-h} plane on a
square and simple cubic lattice for a cluster with two spins is obtained
when the random field is bimodal distributed. \newline
\\
{\it Keywords:} Heisenberg model; Effective Field; Random Field.%
{\newpage}%
\end{abstract}

\section{Introduction}

The random field Ising model (RFIM) has been investigated extensively both
theoretically and experimentally in last years.\cite{shapir92} $^{-}$ \cite
{benayad01} Some questions such as the lower critical dimension and the
existence of a static phase transition have been already solved from the
theoretical point of view, but others questions as the existence of the
tricritical point are still open. However, it is now understood that RFIM
describes the essential physics of a rich class of experimentally accessible
disordered systems. These include structural phase transitions in random
alloys, commensurate charge-density-wave systems with impurity pinning,
binary fluid mixtures in random porous media, and the melting of
intercalates im layered compounds such as TiS$_2$. The behavior of the RFIM
is governed by the competing tendencies of the spins to align
ferromagnetically under impetus of the exchange or to follow local fields
and so be uncorrelated.

The RFIM has been studied with different methods. The renormalization-group
theory (RG) of phase transition and critical phenomena represent one of most
powerful tool of theoretical physics in last decades. Until then (early
1970s) all calculations of critical exponents were made either by exactly or
by series expansions. Therefore, employing the renormalization-group
arguments (which can be applied for weak fields), the proof of the
equivalence between an Ising ferromagnet in a random field and a dilute
antiferromagnet in a uniform field has been demonstrated. These question
have been very explored (see \ref{figueiredo} and references therein).

In the RG scheme, such as, the effective field RG (EFRG)\cite{doug00a}$^{,}$ 
\cite{doug94} and the Mean-field RG (MFRG) \cite{doug97}$^{,}$ \cite{doug00b}
methods have been successfully employed in spin systems, where the results
obtained are in accordance with other more effective approach (series
expansion, Monte Carlo simulation among others). Those approaches (MFRG and
EFRG) are based on comparison of two clusters of different sizes, each of
them simulating infinite systems. On the other hand, the mean field
approximation (MFA) has been extensively applied in the study of RFIM. The
MFA provides very interesting results about RFIM. For example, a bimodal
distribution for random field gives rise to a tricritical point and
consequently first order phase transitions, at very low temperatures for
certain values of field strength, but does not occur in the case of a
gaussian distribution.\cite{aharony78}$^{-}$\cite{galam85} In addition, it
is possible to establish a complete mapping between the parameters of the
Ising ferromagnet in a random field and the dilute Ising antiferromagnet in
a uniform field.

By employing the same strategy of EFRG, the effective field theory (EFT) has
been used to study cooperative phenomena and phase transitions in various
systems and giving useful qualitative and quantitative insights for the
critical behavior of a wide variety of classical and quantum lattice spin
systems.\cite{chakra93}$^{-}$\cite{lima} The approach of EFT uses, as a
starting point, the rigorous Callen-Suzuki spin identities (See Refs.~\ref
{calen63} and~\ref{susuki65}) and the effects of the surrounding spins on
each cluster are taken into account by using a convenient differential
operator technique introduced by Honmura and Kaneyoshi.\cite{honmura79} In
this procedure, all relevant self-spin correlations are taken exactly into
account in the EFT equation of state.\cite{fitti94} Therefore, the EFT
approach is superior to the standard MFA.

In this work, we study the behavior of phase diagram (in $T-h$ plane) for
Heisenberg classical model of spin-1/2 on a square and simple cubic lattice
by employing a bimodal probability distribution for random fields in a
cluster with two spins by using the result recently obtained on diluted
systems.\cite{doug00a} The outline of the remainder of this paper is as
following: in section {\bf 1}, the formalism and calculations are developed
and the results and conclusions are presented in section {\bf 2}.

\section{Formalism and Calculations}

The system to be studied is the $n$-vectorial model in a random field
described by the Hamiltonian 
\begin{equation}
-\beta {\cal H}=K\sum_{(i,j)}{\bf {S}_i}{\cdot }{\bf {S}_j}+\sum_i{{\bf {h}_i%
}}{\cdot }{\bf {S}_i},  \label{eq1}
\end{equation}
where the summation is performed over all pairs of the nearest-neighboring
sites $(i,j)$. The quantities ${\bf {S}_i}$ are isotropically interacting $n$
-dimensional classical spins of magnitude $\sqrt{n}$ localized on site $i$
and the Cartesian components of ${\bf {S}_i}$ obeys the normalization
condition,\cite{stanley69} 
\[
\sum_u^n(S_i^n)^2=n. 
\]
$K$ [$\equiv J/k_BT$, $k_B$ is the Boltzmann constant and T the temperature]
is the exchange interaction between the spins, $h_i$ [$h\equiv \mu _BH/k_BT$
, where $\mu $ is the Bohr magneton and $H$ is the random magnetic field] is
the reduced random magnetic field at site $i$ with probability distribution 
\begin{equation}
P(h_i)=\frac 12[\delta (h_i+h)+\delta (h_i-h)].  \label{eqp1}
\end{equation}

Hamiltonian (\ref{eq1}) reduces to the $S=\frac 12$ Ising, planar (XY),
Heisenberg and spherical models for $n=1,2,3$ and $\infty $, respectively.

In this work, we follow the EFT procedure (see Ref. \ref{rica_a}) to study
the critical properties of the Hamiltonian described by Eq.(\ref{eq1}) by
employing the axial approximation.\cite{idogaki92} Since the Hamiltonian for
a cluster with two spins can be written as

\begin{equation}
H=K{{\bf S}_1}\cdot {{\bf S}_2}+a_1\cdot S{_1^1}+a_2\cdot S{_2^1},
\label{eq2}
\end{equation}
where $a_l=h{_l}+K\Sigma _{j\ne l}^{z-1}S{_j^1}$, ($l=1,2$) and $z$ is the
lattice coordination number.\newline
The average magnetization per-spin $m=\frac 12{\langle \bigl( S_1^1+S_2^1%
\bigr) \rangle }$ for cluster with two spins is given by (See Refs. \ref
{doug0a} and \ref{doug0b}) 
\begin{equation}
m=\langle \prod_{k\ne 1,2}^{z-1}\bigl( \alpha _x+S_k^1\beta _x\bigr) %
\prod_{l\ne 1,2}^{z-1}\bigl(\alpha _y+S_l^1\beta _y\bigr)\rangle
g_n(X,Y)|_{X=h_1,Y=h_2}.  \label{meq2}
\end{equation}
$\alpha _\nu =\cosh (KD_\nu )$, $\beta _\nu =\sinh (KD_\nu )$, $(\nu =x,y),$ 
$X(Y)=x(y)+h_1(h_2)$. $D_\nu $ ($\equiv \frac \partial {\partial \nu }$) is
differential operator which satisfy the mathematical relation 
\[
sinh(aD_x+bD_y)g_n(X,Y)|_{X=h_1,Y=h_2}=g_n(a+h_1,b+h_2),
\]
where $g_n(X,Y)$ is given by 
\begin{equation}
g_n(X,Y)=\frac{\sinh (X+Y)}{\cosh (X+Y)+\exp (-2K)M_n(K)\cosh (X-Y)},
\end{equation}
and 
\[
M_n(K)=\frac{I_{n/2}(nK)-I_{n/2-1}(nK)}{I_{n/2}(nK)+I_{n/2-1}(nK)}.
\]
Here $I_n(X)$ is a modified Bessel function of the first kind. Eq.~(\ref
{meq2}) is exact and will be applied here as a basis of our formalism, since
it yields the cluster magnetization $m$ and the corresponding multi-spin
correlation functions associated with various sites for the cluster under
consideration. Here we apply the EFT approximation in both sides of Eq.~(\ref
{meq2}), i.e., the thermal and random average (denoted by $\langle \cdots
\rangle _c$), along with the decoupling procedure which ignores all
high-order spin correlations, namely $\langle S_i^1S_j^1\cdots S_n^1\rangle
_c\approx \langle S_i^1\rangle _c\langle S_j^1\rangle _c\cdots \langle
S_n^1\rangle _c$ , with $i\ne j\ne \cdots \ne n.$ Based on this
approximation and replacing each boundary configurational spin average by
the symmetry breaking mean-field parameters $b_i$ for all $i$, one set up
the equation of state for $\bar{m}=\langle m\rangle _c$. By using the
properties of differential operator and assuming translational invariance,
we obtain for the square (sq) and simple cubic lattice (sc), respectivelly,

\begin{eqnarray}
\bar{m} &=&\sum\limits_{k=1}^6A_{k,sq}(K,n,h)\bar{m}^k,  \label{meq3} \\
\bar{m} &=&\sum\limits_{k=1}^{10}A_{k,sc}(K,n,h)\bar{m}^k.  \label{meq3a}
\end{eqnarray}
The coefficients $A_{k,(sq,sc)}(K,n,h)$ [$k=$ even] are zero to satisfy the
time reversal symmetry of the Ising model, as well the properties of
differential operator technique.%
\newpage%

\section{Results and Conclusions}

In this section we apply the conditions for determining of the second-order
transition line and the tricritical point (TCP) in the $n-$ vector model in
a random field. We will focus on the case $n=3$, which corresponds to the
Heisenberg classical model of spin-$1/2$. \newline
In the vicinity of the second-order phase transition, the Eqs.~(\ref{meq3} - 
\ref{meq3a}) is given by 
\begin{equation}
\bar{m}^2=-\frac{A_{1,(sq,sc)}(K,n,h)-1}{A_{3,(sq,sc)}(K,n,h)}.  \label{meq5}
\end{equation}
Since the magnetization ($\bar{m}$) goes to zero continuously, to obtain the
second order phase transition line, in the $T-h$ plane, we need solve the
equations 
\begin{equation}
A_{1,(sq,sc)}(K,n,h)=1,\hspace{0.5in}A_{3,(sq,sc)}(K,n,h)<0.  \label{meq4}
\end{equation}
On the other hand, the r.h.s. of Eq.~(\ref{meq5}) must be positive;
otherwise, the transition is interpreted to be of first-order. Unfortunately
the first order transition line cannot to be analyzed in the frame of the
differential operator technique, therefore we have confined our calculations
only to the second order transitions, including the TCP. We obtain the TCP
by solving the equations 
\begin{equation}
A_{1,(sq,sc)}(K,n,h)=1,\hspace{0.5in}A_{3,(sq,sc)}(K,n,h)=0.  \label{meq6}
\end{equation}

The numerical solution of Eq.~(\ref{meq4}) provide the second-order phase
transition line which is shown in Figure 1. When the random field is zero,
we obtain the well-known critical temperature of Heisenberg classical model
of spin-$1/2$, $K_{c,sq}^{-1}=3.005$ and $K_{c,sc}^{-1}=5.031$. In Figure 1,
the solid line indicate the second-order phase transitions; the black
diamonds denote the position of the TCP at which phase transition changes
from second to first order. The numerical solution of Equations (\ref{meq6})
provide the values for the tricritical points $%
(h_{TCP},K_{TCP}^{-1})=(1.800,0.958)$ and $%
(h_{TCP},K_{TCP}^{-1})=(2.785,2.389)$ for the square and simple cubic
lattice, respectivelly. We might conclude saying that EFT approximation is
able to predict the presence of tricritical behavior for the Heisenberg
classical model of spin-$1/2$ on a square and simple cubic lattice in a
random magnetic field which takes on the random values $\pm h$ with equal
probabilities. As can be seen, our results show the existence of the TCP for
square lattice ($z=4$) in accordance with mean-field\cite{rezende} and
Betthe-Peierls\cite{hartz} approximations. This occur because to obtain the
first-order phase transition we must have at least the $\bar{m}^5$ term in
the expansion of Eqs.(\ref{meq3} - \ref{meq3a}). Therefore, in the presente
approach we cannot observe behavior tricritical for lattice with
coordination numbers $z<4$.


\begin{references}
\bibitem{shapir92}  Y. Shapir, in Recent Progress in Random magnets, edited
by D. H. Ryan ( World Scientific, Singapore, 1992), pp. 309-334.

\bibitem{arruda}  A. S. de Arruda and W. Figueiredo, Modern Physics Letters
B, 
{\bf vol. 11}, {\bf nos. 21 and 22}%
(1997) 973.

\bibitem{galam98}  Serge Galam, Carlos S. O. Yokoi and Silvio R, Salinas,
Phys. Rev. B {\bf 57}, 8370 (1998).

\bibitem{benayad01}  N. Benayad, A. Fathi, L. Khaya, Physica A~{\bf 300}
(2001) 225.

\bibitem{reine98}  E. E. Reinehr, W. Figueiredo, Phys. Letters A~{\bf 244}
(1998)165-168.\label{figueiredo}

\bibitem{doug00a}  D. F. de Albuquerque, Physica A~{\bf 287} (2000) 185.%
\label{doug0a}

\bibitem{doug94}  D. F. de Albuquerque and I. P. Fittipaldi, J. Appl. Phys. 
{\bf 75} (1994) 5832.\label{doug94a}

\bibitem{doug97}  D. F. de Albuquerque, I. P. Fittipaldi, J. R. de Sousa,
Phys. Rev. B {\bf 56} (1997) 13650.

\bibitem{doug00b}  D. F. de Albuquerque, J. Magn. Magn. Mater.~{\bf 219}
(2000) 349.\label{doug0b}

\bibitem{aharony78}  A. Aharohy,Phys. Rev. B~{\bf 18} (1978) 3318.

\bibitem{mattis85}  D. C. Mattis, Phys. Rev. Lett.~{\bf 55} (1985) 3009.

\bibitem{kaufman86}  M. Kaufman, P.E.Klunzinger, and A. Khurana, Phys. Rev.
B~{\bf 34} (1986) 4766.

\bibitem{saxena87}  R. M. Sebastianes and V.K.Saxena, Phys. Rev. B {\bf 35}
(1987) 2058.

\bibitem{galam85}  S. Galam, Phys. Rev. B {\bf 31} (1985) 7274.

\bibitem{chakra93}  K. G. Chakraborty, Phys. Lett. A~{\bf 177} (1993) 263.

\bibitem{rica}  J. R. de Sousa and D. F. de Albuquerque, Physica A {\bf 236}
(1997) 419.\label{rica_a}

\bibitem{lima}  A. L. de Lima, B. D. 
Sto\v{s}i\'{c}%
and I. P. Fittipaldi, J. Magn. Magn. Mater. 
{\bf 226-230}%
(2001) 635.

\bibitem{callen63}  H. B. Callen, Phys. Lett.~{\bf 4} (1963) 161.\label%
{calen63}

\bibitem{suzuki65}  M. Suzuki, Phys. Lett.~{\bf 19} (1965) 267.\label%
{susuki65}

\bibitem{honmura79}  R. Honmura, T. Kaneyoshi, J. Phys. C~{\bf 12} (1979)
3979.

\bibitem{fitti94}  I. P. Fittipaldi, J. Magn. Magn. Mater.~{\bf 131} (1994)
43.

\bibitem{idogaki92}  T. Idogaki, N. Uryu, Physica A~{\bf 181}(1992) 173.

\bibitem{stanley69}  H. E. Stanley, Phys. Rev.~{\bf 179} (1969) 570.

\bibitem{rezende}  A. R. King, V. Jaccarino, D. P. Belanger and S. M.
Rezende, Phys. Rev. B.{\bf 32} (1985) 503.\label{rezendea}

\bibitem{hartz}  O. Entin-Wohlman and C. Hartztein, J. Phys. A{\bf 18}
(1985) 315.\label{hartza}
\newpage%
{\Large Figure Captions:} \\ \\
Figure 1. Phase diagram for Heisenberg classical model in presence of the
random magnetic  field, in the $T-h$ plane, on a  square (sq) and simple cubic (sc) lattice.%
\end{references}
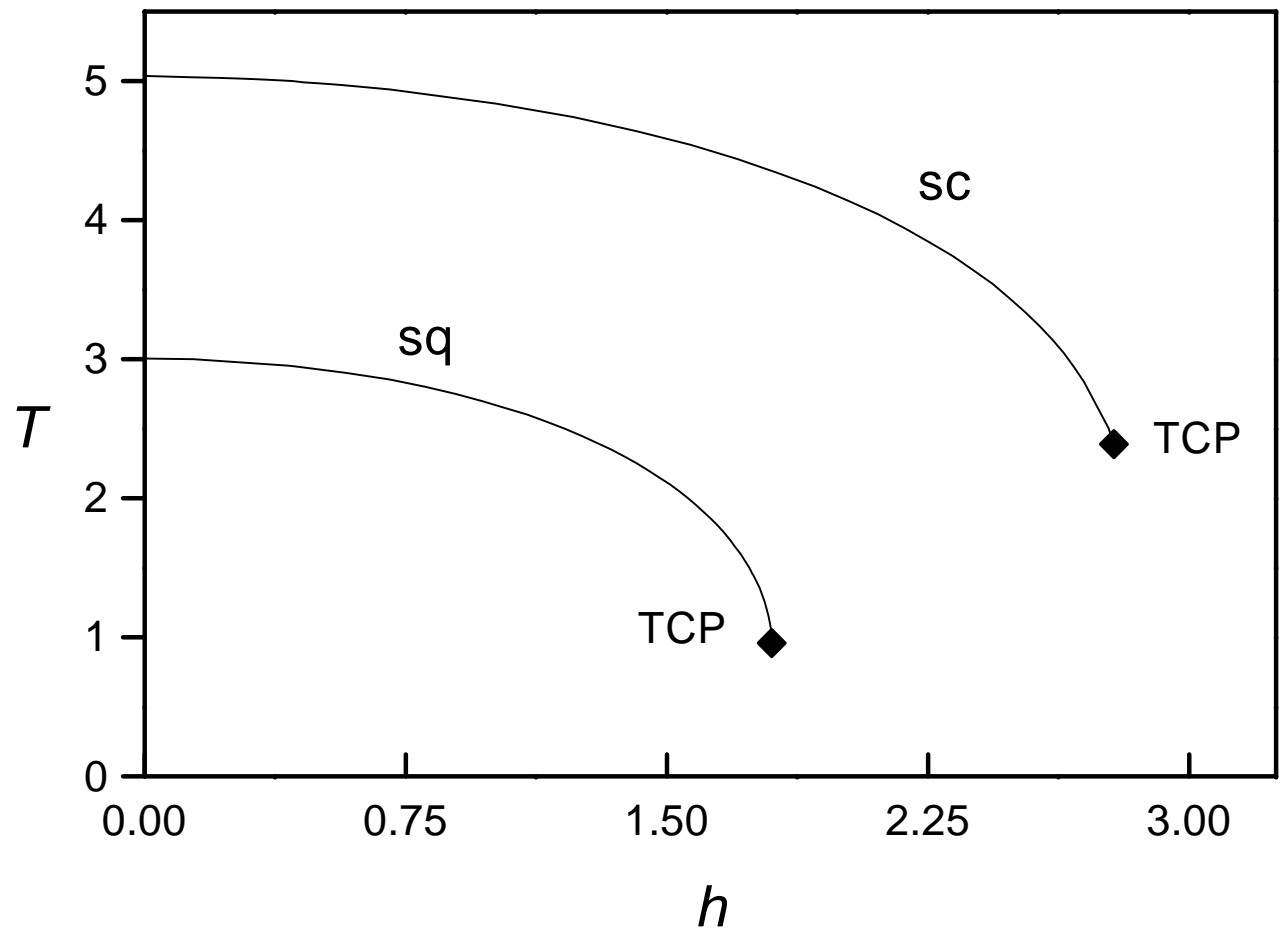
\end{document}